\documentclass[twocolumn,
superscriptaddress,
amsmath,amssymb,aps,showkeys,showpacs,
twoside,final,secnumarabic,%raggedbottom,
nofootinbib]{revtex4-2}

\usepackage[paperwidth=205mm,paperheight=290mm,top=17mm,bottom=25mm,
inner=17mm,outer=17mm,
twoside]{geometry}

\usepackage{cmap} % Улучшенный поиск %русских слов в~полученном pdf-файле
\usepackage[T1,T2A]{fontenc}
\usepackage[utf8]{inputenc}
\usepackage[russian,english]{babel}
\usepackage{color}
\usepackage{graphicx}
\usepackage{epstopdf}% Include figure files
\usepackage{dcolumn}% Align table columns on decimal point
\usepackage{bm} % bold math
\usepackage[unicode=true,colorlinks=true,linkcolor=magenta, urlcolor=blue, citecolor = blue,breaklinks]{hyperref}
\usepackage{multirow}
\usepackage{url}
\usepackage{breakurl}
\DeclareGraphicsExtensions{.eps}

\newcount\issue
\newcount\Vol
\newcount\numb
\usepackage{fancyhdr} %this packages %provides fancy up and bottom of page
\def\Vol{\textbf{80}}
\def\numb{x}
\begin{document}

%====== Начало шапки статьи  ============
\title{JOURNAL SECTION OR CONFERENCE SECTION \\[20pt]
Galactic disc warps from $z = 2.5$ to modern epoch:\\ruling out observational effects} 

\def\addressa{Pulkovo Astronomical Observatory, Russian Academy of Sciences, St.Petersburg 196140, Russia}

\author{\firstname{I.V.}~\surname{Chugunov}}
\email[E-mail: ]{chugunov21@list.ru}
\affiliation{\addressa}
\author{\firstname{V.P.}~\surname{Reshetnikov}}
\affiliation{\addressa}%\affiliation{\addressb}
 \author{\firstname{A.A.}~\surname{Marchuk}}
\affiliation{\addressa}

\received{xx.xx.2025}
\revised{xx.xx.2025}
\accepted{xx.xx.2025}

\begin{abstract}
A significant fraction of galaxies show warps in their discs, usually noticeable at its periphery. The exact origin of this phenomenon is not fully established, although multiple warp formation mechanisms are proposed. In this study, we create a sample of more than 1000 distant ($z \lesssim 2.5$) edge-on galaxies imaged by HST and JWST. For these galaxies, we measurd characteristics of warps and finally analyse how their parameters and frequency change with time. We focus on our main result that galaxies with strong warps were more prevalent in the past compared to the modern epoch. We check how selection effects and varying image quality between objects in our sample could influence our results and conclude that varying fraction of warped galaxies is not caused by observational effects, but represents a genuine evolution. Such a trend may be consistent with mergers and interactions between galaxies being the primary mechanism of warp formation, as number density of galaxies decreases with time, implying higher rate of mergers and interactions in the past.
\end{abstract}

\par
\keywords{galaxies: evolution --- galaxies: high-redshift --- galaxies: spiral --- galaxies: structure   \\[5pt]}
%DOI:  

\maketitle
\thispagestyle{fancy}

%====== Начало  статьи  ============

\section{INTRODUCTION}

Galactic discs, both stellar and gaseous, can demonstrate warps in vertical direction, or, in other words, deviation of their shape from the planar form. The frequency of warps depends on the detection threshold, but common numbers in the literature indicate that nearly half of disc galaxies in the local Universe are warped. The disc of our Milky Way is also long known to be warped~\cite{1, 2}. Warps can be primarily seen in galaxies observed edge-on, they usually have an amplitude of a few degrees and become noticeable at the periphery of a galaxy, making them harder to study~\cite{3, 4, 5}. As such, disc warps at high redshifts are far from being well-studied~\cite{6}.

Although multiple scenarios are proposed to explain the origin and evolution of disc warps, it is still a matter of debate which of them indeed have place. Among discussed ones, there are bending modes in self-gravitating discs~\cite{7}, perturbation of gaseous disc by extragalactic magnetic field~\cite{8}, misalignment of dark matter halo~\cite{9}, cosmic infall~\cite{10}, and, finally, interactions and minor mergers between galaxies~\cite{11, 12}. It is already established that the prominence of warps depends on the environment, with the strongest warps found in the dense environment~\cite{13, 14}. Probably, it can serve as evidence for the importance of the last mechanism.

As such, in this work we study a sample of more than 1000 edge-on galaxies at redshift $z \lesssim 2.5$ and measure the shapes of their discs. In this work, we use a standard flat $\Lambda$CDM cosmology with $\Omega_m$=0.3, $\Omega_{\Lambda}$=0.7, $H_0$=70 km\,s$^{-1}$\,Mpc$^{-1}$, using a calculator of cosmological parameters~\cite{15}.

\section{SAMPLE AND DATA ANALYSIS\label{sec:Methods}}

The final sample of galaxies for this study is presented in~\cite{16} and consists of two subsamples. In the first, there are objects from COSMOS field observed by HST in F814W filter, selected in~\cite{17}. For a small part of galaxies from the mentioned work it was not possible to measure the shape of the disc (see below) and we limited ourselves with galaxies with stellar mass $M_*$ higher than $10^9 M_\odot$, so only 780 galaxies (out of 950 in~\cite{17}) were studied in this work. These galaxies are mostly located at $z \lesssim 1$.

\begin{figure}[h]
\includegraphics[width=0.95\linewidth]{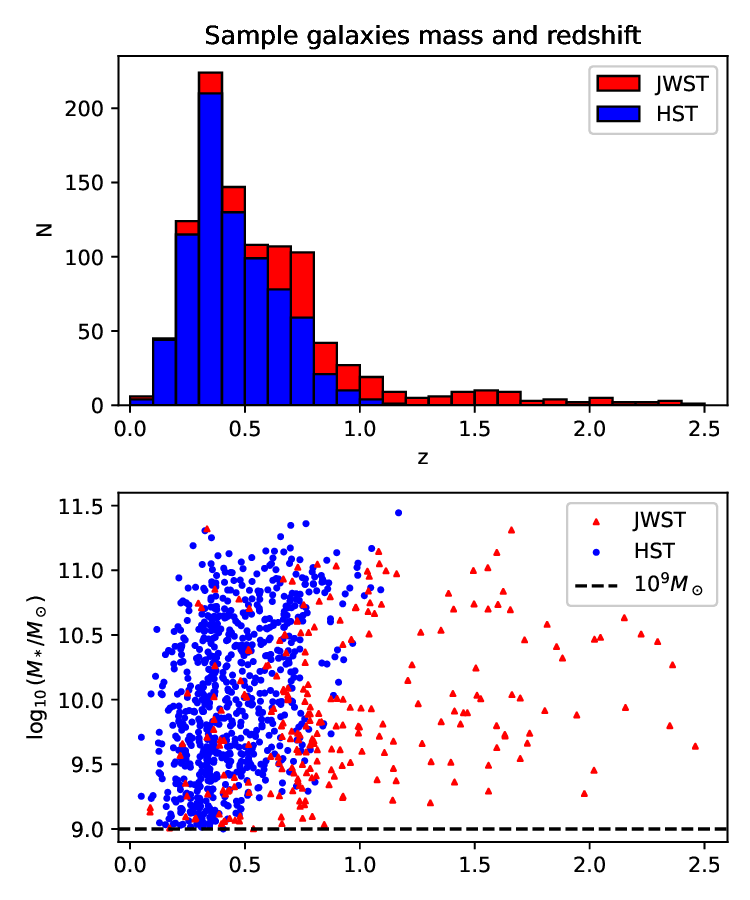}
\caption{\label{fig:zdist} Distribution of galaxies by redshift (top) and the diagram showing stellar mass in solar units and redshift for galaxies in the sample (bottom).}
\end{figure}

The second subsample which was constructed in~\cite{16}, is based on DAWN JWST Archive (DJA) data~\cite{18}, which contains images from several deep JWST fields. For this JWST subsample, we end up with 247 galaxies with images available in F115W or F444W filters, or both. In total, our sample contains 1027 edge-on galaxies up to redshift $z \lesssim 2.5$. Their redshift and mass distribution is shown in Figure~\ref{fig:zdist}. The complete sample is available online at \url{https://github.com/IVChugunov/Distant_disc_warps}.

\begin{figure}[h]
\includegraphics[width=0.95\linewidth]{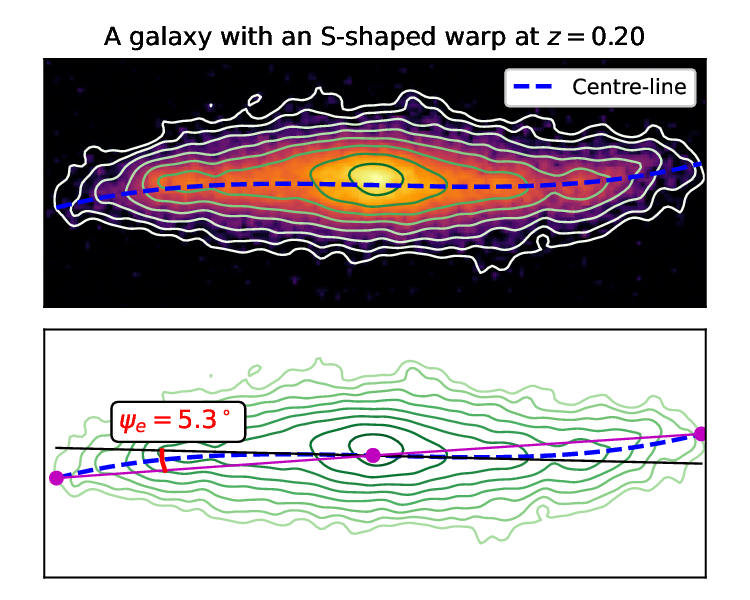}
\caption{\label{fig:skel_example} An example showing a galaxy, its isophotes and measured centre-line (top) and the illustration of how warp angle is defined (bottom). This galaxy is located at $z = 0.20$ and has a conspicuous warp which angle is $\psi_e = 5.3^\circ$.}
\end{figure}

To measure the shapes of discs, in~\cite{16} we used an approach from~\cite{3}. The key method in this approach is the skeletonization of the isophotes. We built multiple isophotes, and, for each one, we selected the area enclosed by it, constructed its skeleton image and selected the longest ``thread'' in it. By definition, skeleton is a one-pixel wide line representing the general shape of the area inside the isophote. Using skeletons of different isophotes, we averaged the locations of skeletons and obtained the centre-line of a disc this way, with an example shown in Figure~\ref{fig:skel_example}. The following measurements of warp properties, in particular their existence in each individual galaxy, warp angle and other parameters, are based on the parameters of centre-line. Formally, centre-line is the set of multiple measurements of the disc offset from the central plane, $\Delta H$, for different galactocentric radii $r$. Central plane is a tangent to the galactic disc in the centre of a galaxy.

Warp angle $\psi_e$ is defined in a manner similar to~\cite{3} and schematically shown in Figure~\ref{fig:skel_example}. However, there is a difference between the definition from this work and from~\cite{3}: we measure $\psi_e$ at the galactocentric radius corresponding to the outermost extent of the available centre-line measurements instead of $R_{25}$. Insufficent image depth, as well as band-shifting effects and cosmological surface brightness dimming make us unable to consistently use $R_{25}$. In the same time, we are fully aware that $\psi_e$ measured in this way inherently depend on the spatial extent of the data, and in Section~\ref{sec:Validation} we pay attention to rule out any possible observational effects.

\section{RESULTS}
For each galaxy in the sample, we measured some of warp properties, most importantly $\psi_e$. We adopt a commonly-used distinction between S-shaped and U-shaped warps based on the apparent shape of the warp (see, for example,~\cite{3}). If a galaxy has warp angle $\psi_e > 4^\circ$, we consider it to have a strong warp. The general results of the measurements were presented in~\cite{16} and now are available online at \url{https://github.com/IVChugunov/Distant_disc_warps}.

\begin{figure}[h]
\includegraphics[width=0.95\linewidth]{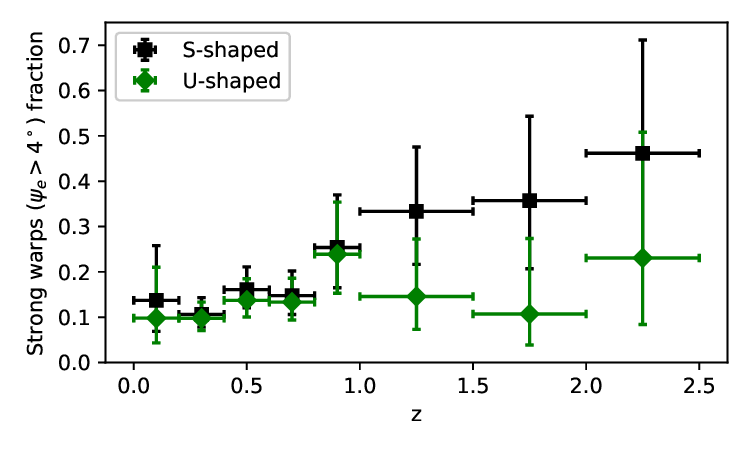}
\caption{\label{fig:warped_frac} This diagram shows the fraction of strong S-shaped or U-shaped warps in different redshift bins. Redshift range of the bin is shown by horizontal error bars, whereas 95\% confidence interval is represented by vertical ones.}
\end{figure}

In Figure~\ref{fig:warped_frac}, the observed fraction of strong warps against redshift is presented. For S-shaped warps, the fraction increases with $z$, from 10--15\% at $z \approx 0$ to nearly 50\% at $z \approx 2$, whereas the fraction of U-shaped warps being around 10--20\%, independently of $z$. The fraction of strong S-shaped warps at low redshift is consistend with literature (for example, 15\% reported in \cite{4} and 11\% in \cite{19}). If one takes galaxies with $\psi_e > 2^\circ$ into consideration, this trend will mostly remain the same: the fraction of S-shaped warps increases with redshift, and for the fraction of U-shaped warps, there will be only a local maximum at $z \approx 1$. We observe no significant correlation between discs $b/a$ where measured by decomposition, and $z$ neither for the entire sample, nor for S-shaped, U-shaped and unwarped galaxies individually. We note that not only warp fraction, but also average warp amplitude moderately increases with redshift (see~\cite{16}).

\subsection{\label{sec:Validation}VALIDATION}
After we observe the trend of warp frequency and amplitude to increase towards higher $z$, we have to ensure that it is not caused by any observational effects and has physical nature. First, our sample is not complete by mass, and galaxies in our data become more massive at higher $z$ (see Figure~\ref{fig:zdist}). Indeed, a connection between the galaxy mass and warp angle exists, but more massive galaxies have smaller warp angles~\cite{3, 10}. Therefore, incompleteness of our sample could not create the observed trend, and the opposite should be expected.

The second possible problem is the varying image quality across the sample, as we noted in Section~\ref{sec:Methods}. Our measurements of $\psi_e$ depend on how far to the periphery we were able to trace the shape of the disc, and for similar galaxies at different redshifts this distance will not be the same. One problem here is band-shifting which arise due to that galaxies at different redshift observed in the same filter are seen in different rest-frame wavelengths. This effect changes the overall surface brightness of a galaxy, which is known as K-correction~\cite{20}. In addition, the dependence of the warp angle on the wavelength itself is also possible~\cite{21}, however in~\cite{16} we compared warp angles between two filters for JWST data, and found no systematic offset. Another effect is cosmological dimming, which manifests itself as a surface brightness decrease for all sources proportional to $(1 + z)^4$~\cite{22}. As a result, we are unable to catch peripheral parts of  the most distant galaxies as they become too faint. It can possibly lead to various effects: for example, at high redshift we may lose the majority of warps which start at large galactocentric distance $z$.

\begin{figure}[h]
\includegraphics[width=0.95\linewidth]{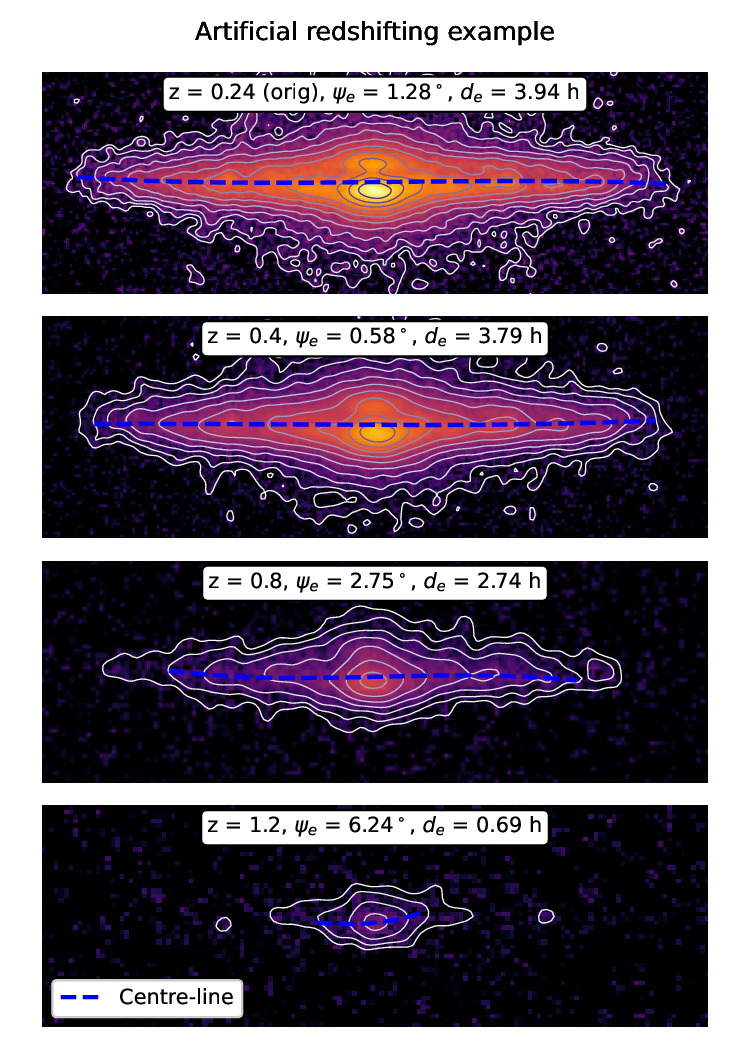}
\caption{\label{fig:artificial_z} An example of a galaxy artificially redshifted to different $z$. Measured warp angles and data extents are also shown.}
\end{figure}

To check this, we performed an artificial redshifting of some galaxies in the sample. Essentially, we decreased image quality in a way to get images how the same galaxy would look at larger redshifts. We used the same approach as in~\cite{23}, reproducing worsened spatial resolution and decreased surface brightness due to K-correction and cosmological dimming. Then, we applied our method of warp measurement to artificial images and we observe examples of different behaviour. In some cases, at large $z$ periphery is lost and one is unable to trace warp anymore; in other cases, as in example in Figure~\ref{fig:artificial_z}, isophotes at large $z$ become too noisy and false warp can appear. However, we did not include too bad images into our sample which makes false warps less likely to appear.

\begin{figure*}[h]
\includegraphics[width=0.95\linewidth]{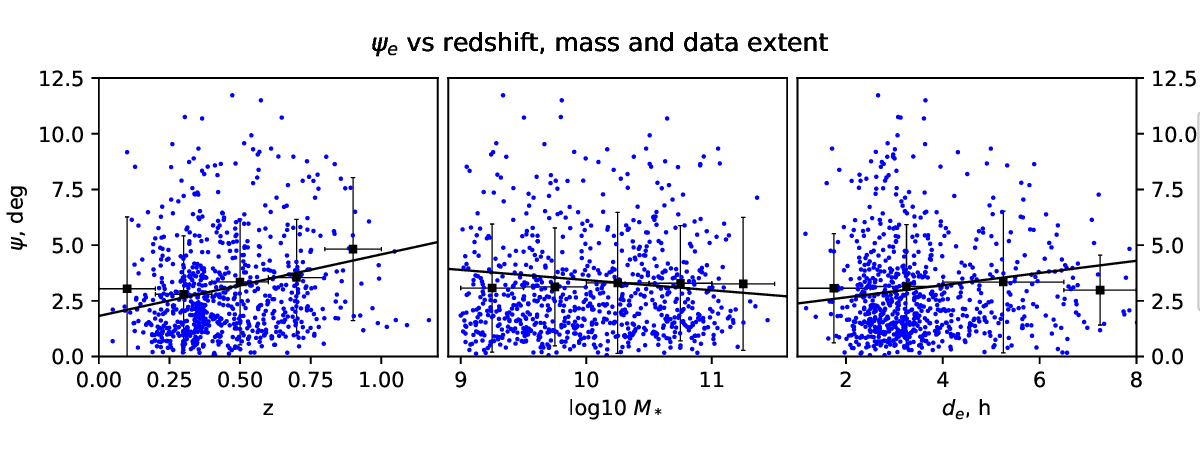}
\caption{\label{fig:tri_lin_warp}The dependence of $\psi_e$ on $z$, stellar mass $M_*$ and data extent $d_e$ and trilinear fit to the data. Black squares with error bars represent value averaged by bins, with the parameter range of the bin being shown by horizontal error bars, and the standard deviation is represented by vertical ones.}
\end{figure*}

In order to measure quantitatively, how strong image quality influences our results, we introduce a parameter of data extent $d_e$. This is the radius at which $\psi_e$ was measured (i.e. the highest galactocentric distance of traced centre-line), normalised to disc exponential scale $h$. Naturally, this parameter represents how far from the centre we can trace warps. Then, we utilise data on $\psi_e$, $z$, stellar mass $M_*$ and data extent $d_e$ for the HST subsample (because reliable $h$ measurements with decomposition are available only for this subsample). We fit $\psi_e$ as a trilinear function of $z$, $M_*$ and $d_e$, aiming to discern the dependencies of $\psi_e$ on these parameters from each other. We show result in Figure~\ref{fig:tri_lin_warp}. Ultimately, we see that $\psi_e$ indeed increases with $z$, even if dependencies on $M_*$ and $d_e$ are taken into consideration, by more than 2.7 degrees per unit of $z$. As expected, the average warp angle decreases with increasing mass and increases with increasing extent of the data, likely representing that if data extent is too small, one risks to miss an existing warp. As a result, we can conclude that the dependence of $\psi_e$ on $z$ is physical, and observational effects are ruled out.

\section{CONCLUSIONS}

In this work, we analysed the sample of more than 1000 distant edge-on galaxies, observed by the HST and JWST. We measured warp angles of their stellar discs and studied how warp frequencies and amplitudes change with redshift. We observe that strong S-shaped warps were much more common in the past, being present in nearly 50\% at $z \approx 2$, whereas at $z \approx 0$ only 10--15\% of galaxies have them. For U-shaped warp, there is no such difference.

We paid attention to check how observational effects could influence our findings. We examined the possible dependence of warp angles on wavelength, galaxy mass and image quality and confirm that our trend is independent on these effects. Therefore, we conclude that S-shaped warp frequency increase with redshift most likely has a physical origin.

This trend can be connected to the increasing frequency of interactions and mergers between galaxies at higher redshift. This is shown by measurements of different indicators of interaction rate, for example statistics of close pairs~\cite{24}, or merger observations~\cite{25}. Therefore, if interactions play a significant role in S-shaped warp formation, one should expect higher fraction of warps in the past, when interactions were more common.

More broadly, the different evolution of S-shaped and U-shaped fractions itself can be evidence of different mechanisms of their formation. For example, tidal interactions are believed to predominantly produce S-shaped warps, and for U-shaped warps, ram pressure stripping is thought to be a possible way to their formation~\cite{5}.

\begin{acknowledgments}
Some of the data products presented herein were retrieved from the Dawn JWST Archive (DJA). DJA is an initiative of the Cosmic Dawn Center (DAWN), which is funded by the Danish National Research Foundation under grant DNRF140.
\end{acknowledgments}

\section*{FUNDING}
This work was supported by the Russian Science Foundation (project No. 24-72-10084).

\section*{CONFLICT OF INTEREST}

The authors of this work declare that they have no conflicts of interest.

%%%%%%%%%%%%%%%%%%%%%%%%%%%%%%%%
% USE thebibliography
%%%%%%%%%%%%%%%%%%%%%%%%%%%%%%%%

\end{document}